\documentclass[12pt]{article}
\usepackage[dvips]{graphicx}
\usepackage{multirow}
\usepackage{epsfig, comment, cancel, ulem, caption, subcaption}
\usepackage{cite}
\usepackage{latexsym,amsmath,amsfonts,amssymb}
\usepackage{mathtools} 
\usepackage{slashed} 
\usepackage{blkarray}
\usepackage[all]{xy} 
\usepackage[latin1]{inputenc}
\usepackage[american]{babel}
\usepackage{bbm}
\usepackage{color}
\usepackage[unicode]{hyperref}
\hypersetup{ linktoc=all,
    colorlinks, linkcolor={palatinateblue},
    citecolor={brightpink}, urlcolor={amaranth}
}

\graphicspath{{Images/}}

\definecolor{rosy}{RGB}{230,235,252}
\definecolor{myframetitle}{RGB}{90,89,170}
\definecolor{myblocktitle}{RGB}{140,185,249}
\definecolor{mytitle}{RGB}{10,80,26}

\definecolor{darkgreen}{RGB}{27,130,45}
\definecolor{darkblue}{rgb}{0,0,0.3}
\definecolor{darkred}{rgb}{0.9,0,0}

\definecolor{light gray}{RGB}{220,220,220}
\definecolor{dark purple}{RGB}{108,0,217}
\definecolor{pink}{RGB}{190,20,100}
\definecolor{orang}{RGB}{193,63,0}
\definecolor{green}{RGB}{11,98,17}
\definecolor{darkpink}{RGB}{153,0,76}
\definecolor{bluegreen}{RGB}{0,102,102}
\definecolor{greenlagan}{RGB}{0,102,0}
\definecolor{redgreen}{RGB}{102,102,0}
\definecolor{Redgreen}{RGB}{153,76,0}
\definecolor{vividviolet}{rgb}{0.62, 0.0, 1.0}
\definecolor{amaranth}{rgb}{0.9, 0.17, 0.31}
\definecolor{palatinateblue}{rgb}{0.15, 0.23, 0.89}
\definecolor{brightpink}{rgb}{1.0, 0.0, 0.5}
\definecolor{cornflowerblue}{rgb}{0.39, 0.58, 0.93}
\definecolor{deepcarminepink}{rgb}{0.94, 0.19, 0.22}
\definecolor{radicalred}{rgb}{1.0, 0.21, 0.37}

\def\6{\partial}

\newcommand{\be}{\begin{equation}}
\newcommand{\ee}{\end{equation}}
\newcommand{\beq}{\begin{equation}}
\newcommand{\eeq}{\end{equation}}
\newcommand{\bea}{\begin{eqnarray}}
\newcommand{\eea}{\end{eqnarray}}

\newcommand{\ba}{\begin{eqnarray}}
\newcommand{\ea}{\end{eqnarray}}

\newcommand{\beqs}{\begin{eqnarray}}
\newcommand{\eeqs}{\end{eqnarray}}
\newcommand{\bal}{\begin{aligned}}
\newcommand{\eal}{\end{aligned}}

\def\lbldef#1#2{\expandafter\gdef\csname #1\endcsname {#2}}

\def\href#1#2{#2}

\newcommand{\ber}{\begin{eqnarray}}
\newcommand{\eer}{\end{eqnarray}}

\newcommand{\beqar}{\begin{eqnarray}}

\newcommand{\eeqar}{\end{eqnarray}}

\newcommand{\dsl}
   {\kern.06em\hbox{\raise.15ex\hbox{$/$}\kern-.56em\hbox{$\partial$}}}

\newcommand{\eeqarr}{\end{eqnarray}}
\newcommand{\ZZ}{{\rm \kern 0.275em Z \kern -0.92em Z}\;}

\def\CC{{\mathchoice
{\rm C\mkern-8mu\vrule height1.45ex depth-.05ex
width.05em\mkern9mu\kern-.05em}
{\rm C\mkern-8mu\vrule height1.45ex depth-.05ex
width.05em\mkern9mu\kern-.05em}
{\rm C\mkern-8mu\vrule height1ex depth-.07ex
width.035em\mkern9mu\kern-.035em}
{\rm C\mkern-8mu\vrule height.65ex depth-.1ex
width.025em\mkern8mu\kern-.025em}}}

\def\RR{{\rm I\kern-1.6pt {\rm R}}}

\def\ZZ{{\rm Z}\kern-3.8pt {\rm Z} \kern2pt}
\def\IB{\relax{\rm I\kern-.18em B}}
\def\ID{\relax{\rm I\kern-.18em D}}
\def\II{\relax{\rm I\kern-.18em I}}
\def\IP{\relax{\rm I\kern-.18em P}}

\newcommand{\bear}{\begin{eqnarray}}
\newcommand{\eear}{\end{eqnarray}}

\def\6{\partial}

\newfont{\namefont}{cmr10}
\newfont{\addfont}{cmti7 scaled 1440}
\newfont{\boldmathfont}{cmbx10}
\newfont{\headfontb}{cmbx10 scaled 1728}

\newcommand{\dd}{\textrm{d}}

\allowdisplaybreaks[1]

\numberwithin{equation}{section}

\begin{document}
\begin{titlepage}

\begin{flushright}

\end{flushright}


 \vskip 1cm

\begin{center}
   \baselineskip=16pt
   {\LARGE \bf A Critique of Holographic Dark Energy
   }
 \vskip 1cm
      {\large Eoin \'O Colg\'ain$^{a, b}$\footnote{ocolgain@gmail.com} \&
      M. M. Sheikh-Jabbari$^c$\footnote{shahin.s.jabbari@gmail.com}
      }

       \vskip .6cm
             \begin{small}
             \textit{$^a$ Center for Quantum Spacetime, Sogang University, Seoul 121-742, Korea}
                 \vspace{3mm}
                 
             \textit{$^b$ Department of Physics, Sogang University, Seoul 121-742, Korea}
                 \vspace{3mm}

   \textit{$^c$  School of Physics, Institute for Research in Fundamental Sciences (IPM),\\ P.O.Box 19395-5531, Tehran, Iran}
                 

             \end{small}
\end{center}


 \begin{center} \textbf{Abstract}\end{center} \begin{quote}
Observations restrict the parameter space of Holographic Dark Energy (HDE) so that a turning point in the Hubble parameter $H(z)$ is inevitable. Concretely, cosmic microwave background (CMB), baryon acoustic oscillations (BAO) and Type Ia supernovae (SNE) data put the turning point in the future, but removing SNE results in an observational turning point at positive redshift. From the perspective of theory, not only does the turning point violate the Null Energy Condition (NEC), but as we argue, it {may be interpreted as an} evolution of the Hubble constant $H_0$ with redshift, which is at odds with the {very FLRW framework within which data has been analysed}. Tellingly, neither of these are problems for the flat $\Lambda$CDM model, and a direct comparison of fits further disfavours HDE relative to flat $\Lambda$CDM. 

\end{quote}
\end{titlepage}
\setcounter{footnote}{0}

\section{Introduction}
A turning point in the Hubble diagram $H(z)$ - concretely a redshift $z_{*}$ where $H'(z_{*}) = 0$ -  is an exotic feature, which is precluded by the Null Energy Condition (NEC) within a Friedmann-Lema\^itre-Robertson-Walker (FLRW) cosmology in Einstein gravity theory. Regardless, cosmological models exist in this class and Holographic Dark Energy (HDE) \cite{Li:2004rb, Wang:2016och} can be regarded  as a prominent minimal model (see also \cite{Huang:2004ai, Pavon:2005yx, Wang:2005jx, Cai:2007us, Gao:2007ep, Chimento:2011dw, Chimento:2011pk, Chimento:2013rya, Tavayef:2018xwx, Saridakis:2020zol} for generalisations). HDE is motivated as a solution to the cosmological constant problem. Constant dark energy density $\rho_{\textrm{de}}$ is not consistent \cite{Cohen:1998zx}  with the holographic principle \cite{tHooft:1993dmi} and consistency with the principle requires  $\rho_{\textrm{de}}$ to have a specific time dependence. Relative to flat $\Lambda$CDM, HDE boasts an additional constant parameter $c$, which current data strongly constrains to the regime $c < 1$,  where the model has turning point; for $0.5 \lesssim c < 1$ the turning point occurs in future  $(z_* < 0)$  \cite{Huang:2004wt, Zhang:2005hs, Chang:2005ph, Zhang:2007sh, Xu:2012aw, Li:2013dha, Zhao:2017urm} while for  $c \lesssim 0.5$, it becomes observational ($z_* > 0$).\footnote{This may be compared with the CPL dark energy model \cite{Chevallier:2000qy, Linder:2002et} with dark energy equation of state $w(z)=w_0+w_a \frac{z}{1+z}$. One may show CPL has a turning point in $z>0$ region if $w_0 < - (1- \Omega_{m0})^{-1}$, where $\Omega_{m0}$ is the matter density.} 

These exotic features are both a blessing and a curse.  On one hand, one may theoretically sidestep the cosmological constant problem, but on the other, it is more difficult to make sense of effective field theories that violate the NEC (however, see \cite{Rubakov:2014jja}).\footnote{NEC violation does not necessarily signal instability in the theory. For example Casimir energy violates NEC. Moreover, ``quantum null energy condition'' (QNEC) has been recently proposed \cite{QNEC} which is a more relaxed condition than NEC and is expected to hold even if there are NEC violations due to quantum effects. So far, to our knowledge no violation of QNEC has been reported. Violation of QNEC, as discussed in the literature, is related to nonunitarity in local quantum field theories.} That being said, HDE may push this to the future, thus postponing it as a pressing concern. Furthermore, a turning point may help observationally with Hubble tension \cite{Verde:2019ivm} - currently one of the biggest puzzles in cosmology - provided the turning point has happened in the recent past $0 \leq z_* \ll 1$. This argument originally appeared in   \cite{mvp2017}, before it was realised in  the HDE model \cite{Dai:2020rfo}. {While presence of a turning point is far from explicit in HDE, it is not an isolated example in the literature, e. g. \cite{DiValentino:2020naf}.\footnote{Other examples may exist and it would be nice to document them.}} Since the turning point in $H(z)$ is at small positive redshift \cite{Colgain:2018wgk}, this overlaps with late-time transition models \cite{Benevento:2020fev}, which as argued in \cite{Camarena:2021jlr} (see also \cite{Alestas:2020zol, Marra:2021fvf}) cannot resolve the Hubble tension. Here, we are not interested in Hubble tension or falsifying turning points, but merely exploring the implications of a turning point within the HDE model.    

HDE has proven itself to be {fairly} adept at mimicking flat $\Lambda$CDM to date {and the idea continues to attract attention largely in theory circles, despite the issues with the NEC violation and presence of a turning point. Nevertheless, if} future data confirms flat $\Lambda$CDM, then HDE will be pushed into a corner and penalised for having an additional parameter. Explicitly put, it won't be able to compete with flat $\Lambda$CDM in a $\chi^2$ comparison. This is a possible outcome (see \cite{Li:2013dha, Zhao:2017urm} for existing constraints) and such $\chi^2$ comparisons resonate well with observational cosmologists. Even if the model is physically well motivated or not, $\chi^2$ provides a handle for a meaningful comparison.  

Here we take a different tack to questioning HDE and pick up a thread explored in \cite{Krishnan:2020vaf}. Concretely, in \cite{Krishnan:2020vaf} it was noted that the Hubble constant $H_0$ is conceptually different than other parameters in a cosmological model; it is an integration constant within the FLRW framework. This latter fact simply follows from the Friedmann equations and is generic to all models once the cosmological principle, i. e. isotropy \& homogeneity at cosmological distances, is assumed. In \cite{Krishnan:2020vaf} it was argued that ``running in $H_0$'' is a sign of the breakdown of the FLRW paradigm within a specific model.\footnote{``Running in $H_0$'' as described in \cite{Krishnan:2020vaf}, alludes to different values one may/will obtain for $H_0$ within a given model of cosmology once data sets at different redshifts are considered.} Here, we recycle this observation {and the intuition gained from it} for HDE.

Our arguments are largely physical and revolve around the inevitability of a turning point in HDE when confronted with the current cosmological data. In particular, the results that guide our insight and support our conclusions are essentially in the literature. To begin, it is worth stressing that the combination CMB \cite{Aghanim:2018eyx}, BAO \cite{Eisenstein:2005su} and SNE \cite{Riess:1998cb, Perlmutter:1998np} have independently confirmed dark energy and the combination CMB+BAO+SNE is completely consistent with the flat $\Lambda$CDM model \cite{Aghanim:2018eyx}. Incidentally, one can find all sorts of interesting tensions in extended $\Lambda$CDM models \cite{DiValentino:2020hov}, but since the extended model is disfavoured relative to the base model, the conclusions drawn elsewhere are irrelevant. Moreover, it is well documented that the combination BAO+SNE provides ``guardrails" at low redshift \cite{Mortsell:2018mfj, Lemos:2018smw, Knox:2019rjx}, which anchor $H_0$ in late-time modifications of flat $\Lambda$CDM to a value consistent with Planck $\Lambda$CDM model value. HDE, being a dark energy model, is no exception and we confirm later that fits to CMB+BAO+SNE data results in a Planck value for $H_0$. Interestingly, removing SNE allows a higher value of the Hubble constant, $H_0 = 71.54 \pm 1.78$ km/s/Mpc \cite{Dai:2020rfo}. As we will verify in this letter, this is due to a turning point in the vicinity of $z = 0$ today.\footnote{In \cite{Dai:2020rfo} $H_0 = 73.12\pm 1.14$ km/s/Mpc and  $c =  0.51 \pm 0.02$ is obtained, but only once a Riess et al. prior \cite{Riess:2019cxk} is included. Still, this figure is illustrative. We provide a similar value without a $H_0$ prior later.}

The picture then is intuitive to anyone with a physics background. Let us spell it out. The combination CMB+BAO+SNE constrains the HDE model to a range of parameter space where it is forced to mimic flat $\Lambda$CDM. Removing the ``guardrails" at low redshift, there is nothing to preclude the turning point moving into the observational regime. Moreover, as BAO data improves with DESI \cite{Aghamousa:2016zmz}, instead of combining CMB with BAO with low weighted redshifts, one can employ BAO with much higher weighted redshifts. In principle, there is nothing to stop the turning point moving to higher redshift, and if this happens, it will lead to even higher values of $H_0$. This is a potential impending issue for the HDE model, but here our analysis will be restricted by the quality of currently available data. Concretely, here we show that adding and removing Pantheon SNE \cite{Scolnic:2017caz} is enough to lead to $\sim 2 \sigma$ displacements in all the cosmological parameters in the HDE model, while $\Lambda$CDM parameters change only by $\sim 0.1\sigma$. Unsurprisingly, these large swings in cosmological parameters lead to jumps in $\chi^2$, but this effect is driven by the unavoidable $H(z)$ turning point in the HDE model. 

Finally, noting that the weighted average redshift for the Pantheon SNE dataset is $z \sim 0.28$, whereas for the employed BAO it is $z \sim 0.36$, one has the basis of a statement that  within the HDE model $H_0$ can run with redshift in the sense discussed in \cite{Krishnan:2020vaf}. {While this evolution of $H_0$ or any cosmological parameter with redshift may sound unfamiliar, note that a valid model is expected to return values of cosmological parameters that are robust to changes in data, i. e. jackknifes. If the data is at different redshifts and there is a mismatch between the model and data, such jackknife exercises can lead to different values at different redshifts or an evolution with redshift.} Recalling that $H_0$ is an integration constant in the Friedmann equations, this contradicts the inherent assumption in the HDE model that it is an FLRW cosmology. Thus, whether one considers i) the NEC violation and the turning point, ii) the running in $H_0$ or iii) a $\chi^2$ comparison, HDE is disfavoured relative to flat $\Lambda$CDM.

\section{Preliminaries}

\subsection{Model review} 
In HDE the basic idea is that the dark energy density takes the form, 
\be
\rho_{\textrm{de}} = 3 c^2 M_{\textrm{pl}}^2 L^{-2}, \quad L := a(t) \int_{t}^{\infty} \frac{\dd t'}{a(t')}, 
\ee
where $M_{\textrm{pl}}$ is the reduced Planck mass and  the length scale $L$ is the future event horizon of the Universe. The latter choice may seem unusual, however, if one assumes simply that $L$ is the Hubble radius (particle horizon), the resulting equation of state (EoS), $\omega_{\textrm{de}}> -\frac13$, does not yield an accelerating Universe \cite{Hsu:2004ri}.

Consider a cosmological model which consists of (dark) matter, radiation and HDE. The relevant equations of motion are, e.g.\cite{Li:2017usw}:
\bea
\label{eq1} \frac{1}{E} \frac{\dd E}{\dd z} &=& - \frac{\Omega_{\textrm{de}}}{(1+z)} \left( \frac{1}{c} \sqrt{\Omega_{\textrm{de}}} + \frac{1}{2} - \frac{3+ \Omega_r}{2 \Omega_{\textrm{de}}} \right) \\
\label{eq2} \frac{\dd \Omega_{\textrm{de}}}{\dd z} &=& -\frac{ (1-\Omega_{\textrm{de}}) \Omega_{\textrm{de}}}{(1+z)} \left( \frac{1}{2c}  \sqrt{\Omega_{\textrm{de}}} + 1 + \frac{\Omega_r}{ 1 - \Omega_{\textrm{de}}}  \right), 
\eea
where $E(z) \equiv H(z)/H_0$ is the normalised Hubble parameter.\footnote{There appears to be a typo in equation (24) of \cite{Li:2017usw}. Flipping the sign in the $\Omega_r$ term in (\ref{eq2}) allows $\Omega_{\textrm{de}}$ to increase at higher redshift.} Note that the algebraic equation
\be
1 = \Omega_m + \Omega_{\textrm{de}} + \Omega_{r}, \quad \Omega_i = \frac{\rho_i}{3 M_{\textrm{pl}}^2  H^2}, 
\ee 
is assumed.  This last condition tells us that $0\leq \Omega_i \leq 1$ in any sector provided the energy densities are positive. As a result, one sees that $\Omega_{\textrm{de}}(z)$ is a monotonically decreasing function of $z$. To take the analysis further, recall that the matter and radiation sectors satisfy the continuity equations $\dot{\rho}_r + 3 H \rho_r = 0$ and $\dot{\rho}_r + 4 H \rho_r = 0$, respectively; here dot stands for derivative w.r.t. comoving time. This fixes $\Omega_i(z)$ in terms of $E(z)$: 
\be
\Omega_m(z) = \Omega_{m0} \frac{(1+z)^3}{E(z)^2}, \quad \Omega_r(z) = \Omega_{r0} \frac{(1+z)^4}{E(z)^2}. 
\ee
$\Omega_{m0}$ is a constant determined through fits to the data and $\Omega_{r0}$ is, in analogous fashion to flat $\Lambda$CDM, fixed by the temperature of the CMB and $N_{\textrm{eff}}$, the number of relativistic (neutrino) species. The equations are solved numerically subject to the conditions that $E(z=0) = 1$ and $\Omega_{\textrm{de}}(z=0) = 1 - \Omega_{m 0} - \Omega_{r0}$. 

As in the flat $\Lambda$CDM cosmological model, radiation is not so relevant at low redshift, and this term can be safely neglected at smaller values of $z$. With $\Omega_r$ removed, it is immediately clear that $ \Omega_{\textrm{de}} $ stops increasing at $ \Omega_{\textrm{de}}  = 1$ in the future as $z \rightarrow -1$ and a de Sitter phase is guaranteed once $c = 1$. Note that the RHS of (\ref{eq1}) and (\ref{eq2}) vanish in this case, so it corresponds to a fixed-point solution for the system.  Moreover, when $c=1$, the HDE model has the same asymptotic attractor as flat $\Lambda$CDM. This can also be seen from the dark energy EoS \footnote{Note that $\Omega_{\textrm{de}}=(c LH)^{-2}$, where $L, H$ are respectively the event and cosmological horizon radii. Therefore, the $\Omega_{\textrm{de}}=1, c=1$ fixed point corresponds to $LH=1$.}
\be
w_{\textrm{de}} = - \frac{1}{3} - \frac{2}{3} \frac{\sqrt{\Omega_{\textrm{de}}}}{c}, 
\ee
which clearly approaches $w_{\textrm{de}} = -1$ as $\Omega_{\textrm{de}} \rightarrow 1$ with $c=1$. Further observing that $\Omega_{\textrm{de}} \approx 0.7$ in the vicinity of $z \sim 0$, we see for $c \lesssim 0.83$ that one encounters a phantom crossing in the observational regime at positive redshift $z$. As we shall see in due course, data places us in a regime with a phantom crossing and hence a de Sitter attractor in the HDE model is disfavoured by data.  

\begin{figure}[htb]
  \centering
  \includegraphics[width=100mm]{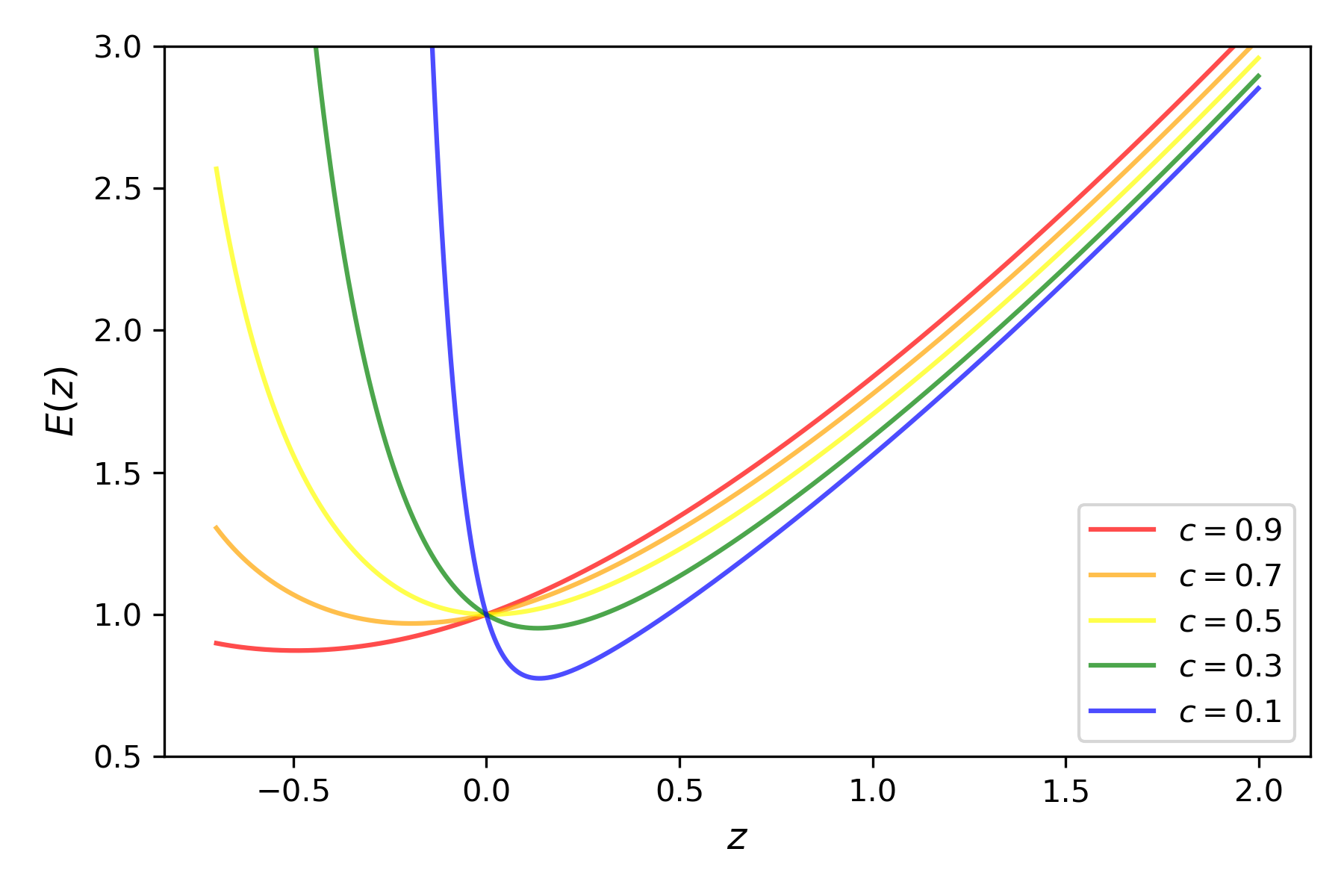}
\caption{Different late-time HDE cosmologies with fixed $E(0) = 1$, $\Omega_{\textrm{de}} (0) = 0.7$ and variable $c < 1$. As $c$ decreases, the turning point moves to higher redshift.}
\label{fig1}
\end{figure}

\subsection{Turning point}\label{sec:2.2}
One interesting feature of the HDE model is the turning point in the Hubble parameter for $c < 1$. Curiously, such an interesting feature was omitted in the review \cite{Wang:2016och}. Observe that even a turning point in the future can influence the Hubble parameter in the past provided it is close enough to $z=0$. Concretely, a future turning point leaves one with a model that largley tracks flat $\Lambda$CDM well at low redshift and it may be difficult to distinguish. However, this conclusion is challenged if data favours $c$ in the regime $c \lesssim 0.5$. 

Recall that at low redshift, where $\Omega_r$ is negligible,  (\ref{eq1})  implies that a turning point happens once 
\be
\frac{1}{c} \sqrt{\Omega_{\textrm{de}}} + \frac{1}{2} \approx \frac{3}{2 \Omega_{\textrm{de}}}. 
\ee
Since $\Omega_{\textrm{de}}<1$ in the observational Universe, this can only happen if $ c < 1$. More precisely, since $\Omega_{\textrm{de}} \approx 0.7$ {is the expected value based on the Planck collaboration's analysis of the $\Lambda$CDM model}, one requires $c \lesssim 0.5$ to have an observational turning point. Conversely, for $0.5 \lesssim c < 1$, there is always a turning point in the future. We document this feature in Figure \ref{fig1} for a range of values of $c$. As can be seen, as $c$ decreases from unity, the turning point crosses over from the future into the past, i.e. within the range of observation. It is worth noting that resolving the Hubble tension with HDE  places us in a regime where the turning point is happening today, $c = 0.51 \pm 0.02$ \cite{Dai:2020rfo}, provided $\Omega_{\textrm{de}} \approx 0.7$. 

\begin{figure}[htb]
  \centering
  \includegraphics[width=120mm]{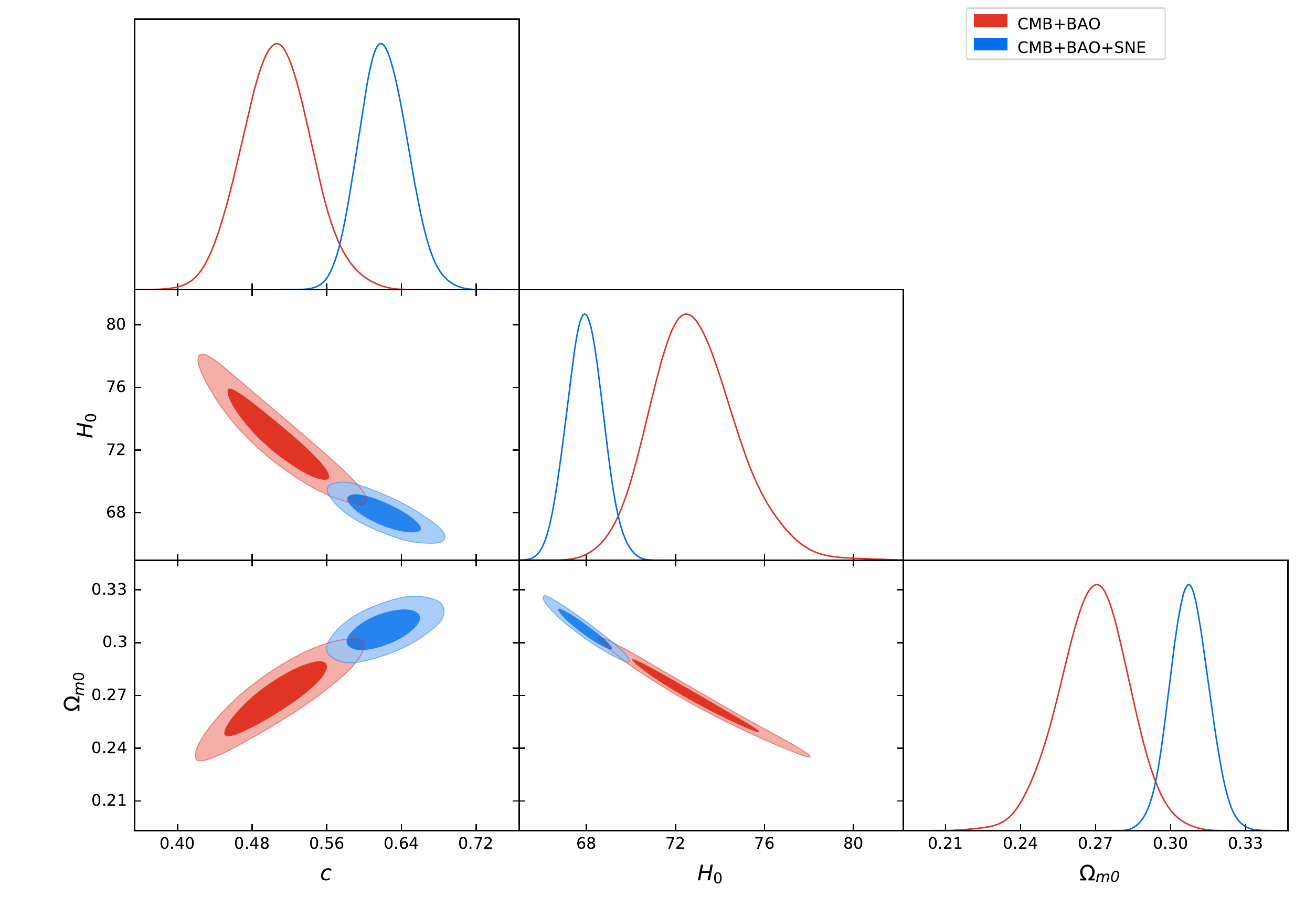}
\caption{The shift in the HDE model parameters following the inclusion of the Pantheon SNE dataset to CMB and BAO. This is to be compared with Figure 3 of \cite{Dai:2020rfo}.}
\label{fig2}
\end{figure}

As is well documented within the context of Hubble tension,  recall again  that the combination SNE+BAO serve as ``guardrails" and confine late-time cosmological models to a $H_0$ value that is close to the Planck-$\Lambda$CDM value \cite{Lemos:2018smw}. Removing the SNE removes the guardrails and allows the HDE model to exploit its natural turning point to raise $H_0$. As is clear from Figure \ref{fig1}, the Hubble parameter is monotonically decreasing in time, just as in flat $\Lambda$CDM, but it starts to increase again after the turning point. This physics is responsible for higher $H_0$ values once the turning point is in the observational regime. This becomes more pronounced when one employs a local prior on $H_0$ \cite{Riess:2019cxk}, but even without, one of the findings of \cite{Dai:2020rfo} is that CMB+BAO data is enough to raise $H_0$. The key point to bear in mind is that HDE  resolves the $H_0$ tension and achieves higher values of $H_0$ (compared to $\Lambda$CDM)  precisely by having a turning point close enough to $z=0$ in the presence of the ``right" data. 

{It should be noted that the current SNE data appears to have some tolerance to a turning point in $H(z)$ at low redshift. To appreciate this, observe that $E(z) \equiv H(z)/H_0$ at $z \sim 0.07$ allows values below unity, $E(z) < 1$ \cite{Riess:2017lxs} within the $1 \sigma$ confidence interval, whereas $E(z=0) = 1$ by definition. In particular, since $E(z)$ can be expanded as $E(z) = 1 + (1+q_0) z + \dots $ at low redshift, this tells one that $1+q_0 < 0$ is permissable in a range of redshift (see also \cite{Camarena:2019moy}), but objectively the data has no preference for it. However, already at $z \sim 0.2$, $E(z) > 1$, so a turning point is only consistent with SNE observations within $1\sigma$ below $z < 0.1$}.\footnote{We thank Adam Riess for discussion on this point.}

Our goal in this letter is to show that this is a double-edged sword: one can find different combinations of reputable data for which the difference in $H_0$ determinations within a given HDE model is sizable. Bluntly put, it is difficult to buy into the notion that $H_0$ is a constant, as required within the FLRW  cosmology framework.

\subsection{Data}
We follow \cite{Dai:2020rfo} and employ the same data. We employ Planck CMB data \cite{Aghanim:2018eyx}, isotropic BAO determinations at  $z = 0.106$ by the 6dF survey \cite{Beutler:2011hx}, SDSS-MGS survey at $z = 0.15$ \cite{Ross:2014qpa} and anisotropic BAO by BOSS-DR12 from $z = 0.38, 0.51$ and $ z = 0.61$ \cite{Alam:2016hwk}. We also use Pantheon SNE \cite{Scolnic:2017caz}. As pointed out in the introduction, the data at redshift lower than $z\sim 0.3$ is dominated by SNE, as the weighted average redshift for the Pantheon dataset is $z \sim 0.28$, and for the  BAO mentioned above, it is $z \sim 0.36$.

\begin{figure}[htb]
	\centering
	\includegraphics[width=120mm]{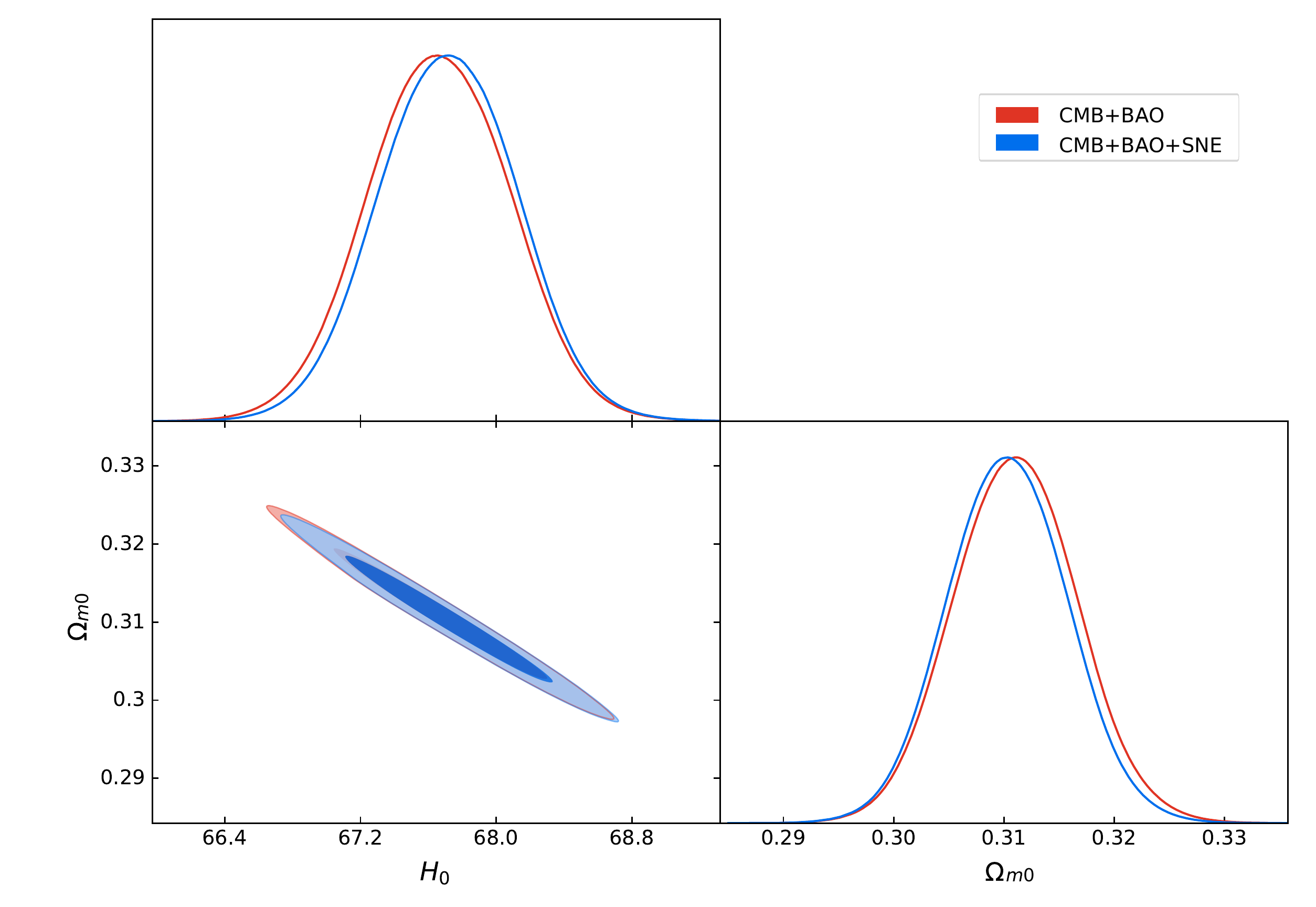}
	\caption{The shift in the $\Lambda$CDM model parameters following the inclusion of the Pantheon SNE dataset to CMB and BAO.  }
	\label{fig3}
\end{figure}

\section{Results}
The main result of our data analysis can be found in Table \ref{Table1}, where we have quoted the best-fit values and the errors obtained from marginalised constraints for both the HDE model and flat $\Lambda$CDM to the data. The corresponding plots can be found in Figures \ref{fig2} and \ref{fig3}. Since, the combination BAO+SNE anchors the Hubble parameter at low redshift \cite{Mortsell:2018mfj, Lemos:2018smw, Knox:2019rjx}, we quote values with and without the Pantheon SNE data. 
Note that we have not employed a local prior on $H_0$ and all our data is cosmological in nature. We ran Markov Chain Monte Carlo (MCMC) for the HDE model, but for flat $\Lambda$CDM we have used  the existing Planck MCMC chains \cite{Aghanim:2018eyx}. The results of our MCMC analysis, while qualitatively similar to \cite{Dai:2020rfo}, differ through the inclusion of dark energy perturbations. 

It is a simple back of the envelope calculation to determine the difference in $H_0$ with and without SNE. We find a discrepancy of $\sim 2.2 \sigma$. This can be contrasted with the analogous number for flat $\Lambda$CDM, namely $0.1 \sigma$, which simply underscores the fact that flat $\Lambda$CDM has an affinity to the data, {and that these three data sets are not mutually inconsistent}. One can confirm from Table \ref{Table1} that this $\gtrsim 2 \sigma$ displacement is not confined to $H_0$ and is also evidently there in both $\Omega_{m0}$ and $c$. This displacement serves as sharp contrast to the flat $\Lambda$CDM model. Since the combination CMB+BAO+SNE is consistent within flat $\Lambda$CDM, it is robust to the addition of SNE data: the addition, as it is intuitively expected, slightly shrinks the confidence ellipses while preserving a significant overlap within $1 \sigma$. This is made explicit in Figure \ref{fig3} as well as in Table \ref{Table1}.

\begin{table}[htb]
	\centering
	\begin{tabular}{c||c|ccc}
		Model & Data   & $H_0$ (km/s/Mpc) & $\Omega_{m0}$ & $c$  \\
		\hline
		\rule{0pt}{3ex}\multirow{2}{*}{HDE } & CMB+BAO & $72.88^{+1.66}_{-2.12}       $ &$0.269\pm 0.014            $& $0.507\pm 0.037            $ \\
		\rule{0pt}{3ex}& CMB+BAO+SNE & $67.94\pm 0.80             $ &$0.308\pm 0.008          $ & $0.621\pm 0.026            $  \\
		\hline 
		\rule{0pt}{3ex}\multirow{2}{*}{$\Lambda$CDM} & CMB+BAO &  $67.66\pm 0.42             $ &  $0.311\pm 0.006          $&  -  \\
		\rule{0pt}{3ex}& CMB+BAO+SNE & $67.72\pm 0.40             $& $0.3104\pm 0.005        $&  - 
	\end{tabular}
	\caption{Best-fit values of the cosmological parameters at 68$\%$C.L.}
	\label{Table1}
\end{table}

The  HDE model is, however,  not robust to the addition of SNE: as depicted in Figure \ref{fig2} and seen in Table \ref{Table1}, there is a clear jump once SNE is added to CMB+BAO. It should be noted that the quoted sigma discrepancies should be treated with caution since there are overlapping datasets, namely CMB+BAO and the actual discrepancy is bounded above by $2.2 \sigma$.\footnote{We thank Yin-Zhe Ma for discussion on this technical point.}  Nevertheless, we believe a like for like comparison between flat $\Lambda$CDM and HDE using the same methodology is meaningful. It is instructive to also record the $\chi^2$ values, which we do in Table \ref{Table2}. Clearly, despite having an additional parameter, HDE fits the data 
 worse than flat $\Lambda$CDM. That point aside, note that the jump in $\chi^2$ when SNE are added is consistent with the introduction of $\sim 1000$ additional data points.

One further observation from Table \ref{Table1} is that for the best-fit values $\Omega_{m0} = 0.308, c = 0.621$ (with SNE), the turning point is located at $z_{*} \approx -0.1$, while for $\Omega_{m0} = 0.269, c = 0.507$ (without SNE), the turning is at $z_{*} \approx 0.04$. This backs up our earlier claim that the turning point is in the observational regime just considering CMB+BAO data alone. 

\begin{table}[htb]
	\centering
	\begin{tabular}{c||c|c}
		Model & Data   & $\chi^2$  \\
		\hline
		\rule{0pt}{3ex}\multirow{2}{*}{HDE } & CMB+BAO & $2818.13$ \\
		\rule{0pt}{3ex}& CMB+BAO+SNE & $3864.86$ \\
		\hline 
		\rule{0pt}{3ex}\multirow{2}{*}{$\Lambda$CDM} & CMB+BAO &  $ 2806.84$  \\
		\rule{0pt}{3ex}& CMB+BAO+SNE & $3841.86$  
	\end{tabular}
	\caption{$\chi^2$ values for different models with different data.}
	\label{Table2}
\end{table}

\section{Discussion} 
Our goal here is not to rule out HDE observationally in the traditional sense using a $\chi^2$ comparison, although as is clear from Table \ref{Table2}, HDE  performs worse than flat $\Lambda$CDM despite having one more parameter. 
Nevertheless, as we do here, one can in tandem comment on the theoretical assumptions going into the HDE model and whether they are borne out in observations. Based on vanilla cosmological data, namely CMB, BAO and SNE, we find evidence for $\sim 2 \sigma$ running in cosmological parameters, especially in $H_0$. Such a feature is not evident in flat $\Lambda$CDM, so HDE, {unlike flat $\Lambda$CDM}, is in conflict with some component of the data and robustness appears to be a problem. {This feature and the higher $\chi^2$ can be traced to the turning point in $H(z)$, which itself is a blatant violation of the NEC that places HDE at odds with quantum physics. One of our objectives here is to draw attention to} the turning point in $H(z)$, which has not received due attention in observational cosmology studies of the model. Remarkably, it fails to feature in the review \cite{Wang:2016och}. 

Within the cosmological parameters of the HDE model, we single out $H_0$ as being special on the grounds that it is an integration constant in the Friedmann equations, and thus common to \textit{all} FLRW cosmologies. 
Our results, based on the current data, are in noticeable tension with the idea that $H_0$ is a constant.\footnote{We are conscious that this tension with FLRW may not be a problem in the end. It is imperative to test the cosmological principle, e. g. \cite{Secrest:2020has, Siewert:2020krp}. See \cite{Cai:2021wgv} for a resolution of the Hubble tension where FLRW is relaxed.} To appreciate this, observe that for $c < 1$ a turning point in $H(z)$ within HDE is unavoidable. The overall combination CMB+BAO+SNE imposes strong enough constraints that HDE is forced to mimic flat $\Lambda$CDM and this exorcises the theoretically unsavoury turning point to the future. {In some sense, a future turning point still allows one to treat HDE as a consistent effective quantum theory in the observational regime.} However, removing low redshift data, in particular the Pantheon SNE, allows the turning point to return to the past and this results in a higher $H_0$ that is discrepant at the $\sim 2 \sigma$ level with the value fixed by CMB+BAO+SNE.\footnote{As mentioned in the second last paragraph of section \ref{sec:2.2}, SNE data disfavours  a turning point in $H(z)$ in the $z<0.1$ region. This is consistent and confirms the statement above.} While admittedly the $\sim 2 \sigma$ discrepancy is an overestimation, because there is overlapping data, the same overestimation logic applies to the $\sim 0.1 \sigma$ discrepancy seen in the flat $\Lambda$CDM model. Clearly, the like for like comparison has meaning. As can be argued, the effective SNE redshift is lower ($z \sim 0.28$) than the effective BAO redshift ($z \sim 0.36$), so we are seeing preliminary evidence for a running $H_0$ with the redshift of the data, as explained in \cite{Krishnan:2020vaf}. We expect similar conclusions to hold for generalisations of the HDE paradigm.\footnote{The model presented in \cite{mvp2017} has one parameter less than HDE and it is enough to add or remove  Lyman-$\alpha$ BAO \cite{Agathe:2019vsu, Blomqvist:2019rah} determinations of the Hubble parameter (instead of the SNE) to find displacements in $H_0$. Simply put, if Lyman-$\alpha$ BAO holds up, then the HDE model presented in \cite{mvp2017} can be falsified.} Upcoming DESI releases will provide better quality BAO to much higher redshifts, which may permit the turning point in $H(z)$ to venture deeper into the past, resulting in even higher $H_0$ inferences. This can be investigated through comprehensive jackknifes of the forthcoming data. This ``running $H_0$", if substantiated, may provide a means to rule out the HDE model, and potentially related models, without resorting to a $\chi^2$ comparison.

Note that our conclusions can be squared with other results in the literature, in particular Figure 3 of Dai et al. \cite{Dai:2020rfo}. Tellingly, the grey contours CMB+BAO+SNE are consistent with Planck. Removing the Pantheon SNE for $z < 0.2$,  in addition to  removing them completely, leads to the blue and green contours, respectively, and the resultant higher values of $H_0$. This is consistent with a turning point at positive redshift. The only contour that is mysterious is the red contour. But here again, there is an interesting explanation. It is well documented that Pantheon prefers a lower value of $\Omega_{m0}$ (effectively the deceleration parameter) below $z \sim 0.2$ \cite{Colgain:2019pck,mvp2020, Camarena:2019moy}. Furthermore, as is clear from Figure \ref{fig3}, the HDE model shares the same degeneracy in the parameters $(H_0, \Omega_{m0})$ as flat $\Lambda$CDM. This means that as $\Omega_{m0}$ goes down (and it goes down in Pantheon for $z \lesssim 0.2$!) the SNE can tolerate a higher value of $H_0$ driven by CMB+BAO data. Consideration of a $H_0$ prior \cite{Riess:2019cxk} only makes this trend, which is driven by the turning point, more pronounced. So, yes, HDE can alleviate Hubble tension, but at the relatively high price of violating the underlying FLRW assumption that $H_0$ is a constant {and that its inferred value is robust under inclusion/removal of the SNE data set}. This is in addition to violating the NEC and presence of a turning point in $H(z)$. This is essentially a clash between theory, or model assumptions, and observation.  

{Finally, let us take the opportunity to comment on various generalisations of the minimal HDE model \cite{Huang:2004ai, Pavon:2005yx, Wang:2005jx, Cai:2007us, Gao:2007ep, Chimento:2011dw, Chimento:2011pk, Chimento:2013rya, Tavayef:2018xwx, Saridakis:2020zol} and their status when compared to observation. One can add spacetime curvature \cite{Huang:2004ai}, but the combination CMB+BAO+SNE is consistent with a flat Universe, so this generalisation is not well motivated. As explained in the review \cite{Wang:2016och}, Agegraphic Dark Energy \cite{Cai:2007us} is ``strongly disfavored" by the data.  HDE based on Ricci curvature \cite{Gao:2007ep} violates the NEC as can be seen from existing observational constraints \cite{Zhang:2009un}.  Big Bang Nucleosynthesis (BBN) places strong constraints on the additional parameter in Barrow HDE, $\Delta  \lesssim 1.4  \times 10^{-4}$ \cite{Barrow:2020kug}, so much so that the model reduces to minimal HDE and either the results of \cite{Hsu:2004ri} or our analysis here applies. There is evidently a zoo of possibilities beyond the minimal HDE model, but one imagines that each iteration of the idea is problematic at some level.}

\section*{Acknowledgements}
We thank Stephen Appleby, Qing-Guo Huang, Yin-Zhe Ma, Adam Riess and Xin Zhang for discussion and/or comments on earlier drafts. E\'OC is funded by the National Research Foundation of Korea (NRF-2020R1A2C1102899). MMShJ acknowledges SarAmadan grant No. ISEF/M/400122. 

We are also especially indebted to Tao Yang and Lu Yin for discussions, running MCMC analysis, analysing the MCMC chains and producing plots. 


\end{document}